# Observation of (N$_2$)$_2$ dimers in free nitrogen and argon-nitrogen clusters

Yu. S. Doronin, M. Yu. Libin, V. N. Samovarov, and V. L. Vakula[*]

B. Verkin Institute for Low Temperature Physics and Engineering

of the National Academy of Sciences of Ukraine,

47 Lenin Avenue, Kharkiv, 61103, Ukraine

## Abstract

Supersonic-jet luminescence spectroscopy was applied to study vibronic transitions in icosahedral N$_2$ and Ar-N$_2$ clusters having from 100 to 400 particles per cluster. In the case of mixed Ar-N$_2$ clusters, the $w^1\Delta_u \to X^1\Sigma_g^+$ transitions were observed to occur in single N$_2$ molecules in an Ar environment, that is very much in the same way as in Ar-N$_2$ bulk samples. In N$_2$ clusters, however, a band series was detected which was never observed earlier. In the spectra of Ar-N$_2$ clusters, this new series coexisted with the 'bulk' $w^1\Delta_u \to X^1\Sigma_g^+$ transitions. Our analysis demonstrated that the series should be assigned to emission of van der Waals (N$_2$)$_2$ dimers from inside clusters. Earlier, such dimers were only observed in molecular beams and gaseous nitrogen, this paper reports their observation in the solid phase of nitrogen suggesting that they may be the basic building blocks of small N$_2$ clusters. Our results can be of interest from the viewpoint of producing polymeric nitrogen since (N$_2$)$_2$ dimers can be considered to be a starting species for its synthesis.

# I. INTRODUCTION

Despite the extensive experimental research on physical properties of bulk nitrogen in the different phases in a large temperature range, a very limited amount of data is available on nitrogen clusters. The experimental studies were mainly concerned with the cluster structure, and only few papers were published on spectroscopy of electronic and vibrational transitions of molecular nitrogen in clusters. As an example, we can mention the Raman spectra from nitrogen clusters in the β-solid and α-solid phases[1], which are typical of bulk crystals, reported in Ref. [3] (see also earlier references on Raman spectroscopy cited therein). Photoelectron spectroscopy and mass spectroscopy were applied to study small $(N_2)_n$ clusters ($n \leq 10$) in charged and neutral molecular beams, including supersonic beams, in order to find the cluster ionization energy and to detect the $(N_2)_2^+$ and $(N_2)_3^+$ ions, as well as tetranitrogen ($N_4$) species [4-6].

The studies of larger atomic and molecular clusters were concentrated mostly on size effects, especially in the vicinity of the transformation from the icosahedral structure, not observed in bulk cryocrystals, to a 'bulk' crystalline structure. Such a structural transformation is quite usual in free intermediate-size clusters, i.e. those containing from a few tens to a few thousands of particles. Smaller intermediate-size clusters start with a polytetrahedral (amorphous) structure, which transforms into a multilayer icosahedral structure when the number of particles exceeds 100, and then to a crystalline cubic structure in clusters with 1500-2000 particles; as clusters grow further, a hexagonal phase arises to coexist with the cubic structure [1,2,7]. There are no sharp transitions in either one- or two-component clusters from one structure to another similar to those observed in bulk samples, since several size-dependent structures (e.g., icosahedral and fcc, icosahedral and hcp) can actually coexist in the cluster beam in a certain range of cluster size [8].

---

[1] According to the electron diffraction data, the cubic α phase is observed in nitrogen clusters with about 4000 molecules [1], while the pure β (hcp) phase is pronounced in clusters containing at least $5 \cdot 10^4$ molecules [2].

Electronic spectra of atoms, polytetrahedral and icosahedral van der Waals clusters, and microcrystals of rare gases have been extensively studied experimentally [9-11], but, to our knowledge, no similar experiments have been performed with nitrogen.

The present paper presents the cathodoluminescence spectra of free $N_2$ and Ar-$N_2$ clusters produced a free supersonic jet expanding into vacuum. Our measurements were performed on: (i) beams of polytetrahedral $N_2$ clusters containing, on the average, about 100 molecules per cluster and (ii) beams of multilayer icosahedral Ar-$N_2$ clusters with the average number of particles of about 250 and 400.

The identified part of the emission spectrum from $N_2$ clusters was found to be made up solely of electronic transitions occurring to vibrational levels of the ground state of $(N_2)_2$ van der Waals dimer. In Ar-$N_2$ clusters, these transitions were accompanied by emission bands of single $N_2$ molecules. The dimers have not been observed in bulk solid samples, but only in gaseous nitrogen [12,13] and molecular beams of nitrogen [4,5].

Tetranitrogen species can exist not only as van der Waals dimers, but also as molecular $N_4$ isomers [6,14,15], the lowest-energy bound isomer being the triplet $N_4$ in $C_s$ symmetry [15]. A tetranitrogen molecule was first observed in the experiments based on neutralization-reionization mass spectrometry of gaseous nitrogen [14], the symmetry of the molecule was shown to be $N_4(C_s)$ [6]. It was demonstrated both experimentally and theoretically that all of the isomers are metastable: for example, the lifetime of the $N_4(C_s)$ isomer was found to exceed 1 μs [6,14]. Some of the isomers are found to collapse into two nitrogen molecules [15], while others can converge into the T-shaped van der Waals complexes $(N_2)_2$ [6].

In recent years, considerable interest in such small nitrogen species has been stimulated by the search for high-energy-density (HED) materials in polynitrogen compounds, particularly in polymeric (single-bonded) nitrogen. Some spectroscopic evidence for single-bonded monoatomic phase of nitrogen was obtained in experiments with solid nitrogen compressed up to 100 GPa [16-18]. There are theoretical grounds for believing that polymeric nitrogen may be metastable at atmospheric pressure [18].

Single bonds, responsible for the great interest in nitrogen compounds as potential HED materials, can also be formed in small nitrogen complexes such as tetranitrogen [15]. There is some experimental evidence for the formation of polymeric nitrogen clusters ($N_4$ and $N_8$) on sidewalls of carbon nanotubes under ultraviolet irradiation [19]. Our solid-state observation of $(N_2)_2$ dimers, which can be a starting species for the formation of single-bonded $N_4$ isomers supposedly resulting from excited states of nitrogen [15], can motivate experimental study on the possibility of accumulating the isomers by exciting nitrogen clusters.

## II. EXPERIMENT

The cathodoluminescence experiments were carried out on free nitrogen and argon-nitrogen clusters formed in a supersonic gas jet flowing into vacuum. The method was analogous to the one used in our study of exciton-impurity complexes in mixed Xe-Ar clusters [11]. A cluster beam was produced by a metallic conical nozzle with the throat diameter $d = 340$ μm and the cone opening angle $2\alpha = 8.6°$. The vacuum chamber was evacuated by a condensation pump cooled with liquid hydrogen. The pressure $p_0$ and the temperature $T_0$ of the primary gas at the entrance to the nozzle were 1.5 atm and 210 K (nitrogen); 1.5, 2 atm and 170 K (argon-nitrogen mixture). Pure nitrogen and mixed argon-nitrogen cluster beams were probed at a distance of 30 mm from the nozzle exit where their temperature was virtually constant, being about $T_{cl} \approx 40$ K. The argon-nitrogen gas mixture contained 5 at.% of nitrogen and, according to Ref. [8], the nitrogen concentration in clusters was approximately the same value. It should be noted that the Ar-N$_2$ solid system is a substitutional solution over the whole range of impurity concentration[2].

The molecule number in the clusters was estimated by using the so-called 'Hagena parameter' [21]:

$$\Gamma^* = k \left( \frac{c \cdot d(\mu m)}{\tan \alpha} \right)^{0.85} \frac{P_0(mbar)}{T_0^{2.29}}, \quad (1)$$

where the parameters $k = 528$ [21] and $c = 0.866$ [22] refer to nitrogen. The average number of molecules in a cluster can be calculated from the following relation based on the electron diffraction data on nitrogen clusters [2]:

$$N = 5 \cdot 10^{-6} (\Gamma^*)^2. \quad (2)$$

According to Eq. (2), the number of molecules in the pure N$_2$ clusters was $N \approx 100$ ($p_0 = 1500$ mbar, $T_0 = 210$ K), while the mixed Ar-N$_2$ clusters contained $N \approx 250$ ($p_0 = 1500$ mbar, $T_0 = 170$ K) and $N \approx 400$ particles ($p_0 = 2000$ mbar, $T_0 = 170$ K). The clusters with $N \approx 100$ were of a polytetrahedral structure made up of randomly oriented elementary icosahedra, each having 13 molecules and 20 faces. The clusters with $N \approx 250$ and $N \approx 400$ particles were multilayer icosahedra (quasi-crystals). A multilayer icosahedron consists of 20 identical tetrahedra with a common vertex in the centre of the inner elementary icosahedron [23].

The cathodoluminescence spectra were measured in the region 45000-73000 cm$^{-1}$ (5.6-9.1 eV), where several molecular emission bands of nitrogen were expected to be observed (these bands are discussed in the next section). The electron energy (1 keV) and beam current (20 mA) were kept constant in all of our measurements in order to enable comparisons between the spectra obtained for different parameters of the primary gas mixture. The measurements were made by using a vacuum monochromator with a spectral resolution of 30 cm$^{-1}$ at the low-frequency end near 45000 cm$^{-1}$ and of 90 cm$^{-1}$ at the high-frequency end near 73000 cm$^{-1}$; these values are smaller than the characteristic half-widths of the studied spectral features. A series of experiments was carried out with various scanning pitches not exceeding the spectral resolution. This technique was successfully applied to study luminescence spectra from pure and mixed rare-gas clusters (e.g., exciton-impurity luminescence features [11]), phase segregation and emission from neutral and charged excimer molecules in mixed Ar-Xe clusters [24], it has also been used in the studies of polarization bremsstrahlung arising from electron scattering on Xe clusters performed on a similar experimental setup [25].

---

[2] In Ar-N$_2$ clusters with no more than 100 particles, a phase segregation occurs, resulting in the formation of an argon core [20].

### III. RESULTS

Figure 1a shows a general cathodoluminescence spectrum from pure nitrogen clusters ($\delta \approx 20$ Å, $N \approx 100$ molecules/cluster) in the range 50000-70000 cm$^{-1}$. It has a complicated structure with a great number of spectral features varying strongly in intensity. It seems quite natural to start analyzing the spectrum by comparing it with the spectra of bulk nitrogen samples that were studied in detail previously. Figure 1b shows the cathodoluminescence spectrum of a 100-μm thick polycrystalline sample of nitrogen with crystal grains of more than 1000 Å in size measured in the same frequency range at 5 K after excitation of the sample by electrons with the energies of 0.5 keV [26] (see also Ref. [27]). The bulk sample's spectrum is rather easy to analyze. In our frequency range, it contains three well-known vibronic (electronic-vibrational) band series of N$_2$ arising from transitions between the vibrational levels of an excited electronic state and the ground electronic state: $a'^1\Sigma_u^- \rightarrow X^1\Sigma_g^+$ (the energy of the transition between the zero vibrational levels of the two states in a free molecule is $\nu_{00} = 67739.3$ cm$^{-1}$), $a^1\Pi_g \to X^1\Sigma_g^+$ ($\nu_{00} = 68951.2$ cm$^{-1}$), and $w^1\Delta_u \to X^1\Sigma_g^+$ ($\nu_{00} = 71698.4$ cm$^{-1}$). In the solid phase, the three series are shifted toward the red end with respect to the free molecule spectrum by 510 cm$^{-1}$ ($a'$, $a \to X$) and 520 cm$^{-1}$ ($w \to X$). The vibrational frequencies and anharmonicities of the excited and ground states remain virtually unchanged compared with those of free molecules which significantly facilitates the identification of transitions. The most pronounced emission band series in Fig. 1b corresponds to the $a^1\Pi_g \to X^1\Sigma_g^+$ transitions. However, we will see it below, the most interesting series for the clusters is the $w^1\Delta_u \to X^1\Sigma_g^+$ one. In Fig. 1b are shown the transitions of this series from the lowest vibrational level of the excited electronic state w ($v' = 0$) to the various vibrational levels of the ground electronic state X ($v'' = 2, 3,\ldots, 7$).

We would like to note that the vibronic lines in crystalline films of nitrogen are noticeably broader than those observed in the cluster spectra (cf. Figs. 1b and 1a). In a nitrogen crystal with the unit cell having two N$_2$ molecules, this broadening is due mainly to the Davydov splitting which may be as great as 100-150 cm$^{-1}$ [27]. In non-crystalline media, such as icosahedral clusters, the Davydov splitting is absent and the vibronic lines are narrower.

The emission spectrum from clusters appears much more complicated, especially in its low-frequency part, the number of spectral features being nearly twice that for bulk samples. In spite of its rich structure, we failed to find the electron-vibrational series corresponding to those observed in the emission spectra from bulk samples of nitrogen.

Figure 2 shows more detailed emission spectra from pure nitrogen clusters in three frequency ranges (45000-73000 cm$^{-1}$). Emission lines of neutral and ionized atomic nitrogen are marked as N and N$^+$. The lines at 46753 cm$^{-1}$ (N$^+$), 54204 cm$^{-1}$ (N$^+$), and 57303 cm$^{-1}$ (N) are only slightly different in their energy positions from the emission lines of free nitrogen atoms and N$^+$ ions. They can be assigned to the emission of atoms and ions desorbed from clusters into the vacuum as well as to the weak emission of the residual non-condensed gaseous component in the cluster beam[3]. The other three emission lines at 63005 cm$^{-1}$ (N$^+$), 67103 cm$^{-1}$ (N), and 71025 cm$^{-1}$ (N) are shifted by about 100 cm$^{-1}$ toward the red end, which implies that the emitting atoms and ions are within the cluster[4]. We'd like to note that the intensity of the ionic line at 63005 cm$^{-1}$ is rather high. The observed emission of ionic centers in rare gas clusters is also very intense [10], since there is virtually no recombination of the ion core with the ionized electron possessing enough kinetic energy to leave the cluster.

The other spectral features in Fig. 2 should be attributed to the emission from molecular states of nitrogen. We could expect a somewhat simpler spectral pattern from mixed Ar-N$_2$ clusters with N$_2$ serving as impurity. Additional measurements on such mixed Ar-N$_2$ clusters allowed us to register one of the electronic-vibrational series observed in the emission spectra from bulk samples of pure nitrogen.

### A. Luminescence of single N$_2$ molecules from Ar-N$_2$ clusters

Figure 3 shows a section of the emission spectrum from Ar+5%N$_2$ clusters ($\delta \approx 30$ Å, $N \approx 400$ particles/cluster) with the $w^1\Delta_u \to X^1\Sigma_g^+$ electronic-vibrational transitions (indicated by vertical solid lines; v′ = 0; v″ = 2, 3,…,7) in N$_2$ molecules isolated in an Ar matrix (cf. with Fig. 1b). It can be seen that the intensity distribution is nearly the same as in the spectra from N$_2$ molecules in nitrogen bulk samples. It is convenient to analyze the spectra from bulk samples and clusters by comparing the energy positions of the lines of a band series $E_{cl}(v′,v″)$ with those from free molecules $E_{gas}(v′,v″)$. For a free molecule in a gas, the quantization of electronic-vibrational levels is given by the following relations:

$$\varepsilon(v′) = \hbar\omega′(v′ + 1/2) - \hbar\omega′x′(v′ + 1/2)^2 \qquad \text{for an excited electronic state,} \qquad (3)$$

---
[3] The spectral contribution of gaseous nitrogen is rather small since the pressure of the non-condensed component is nearly 0.1 Torr in the excitation zone.
[4] A similar 100-cm$^{-1}$ red shift of the atomic nitrogen emission lines was registered in the visible region of bulk nitrogen spectra in Ref. [28].

$$\varepsilon(v'') = \hbar\omega''(v''+1/2) - \hbar\omega''x''(v''+1/2)^2 \qquad \text{for the ground electronic state,} \qquad (4)$$

here $\omega$ and $\omega x$ are the vibrational frequencies and anharmonicities of the two states. These relations with slightly different frequencies and anharmonicities can also be used to describe vibrations of molecules in a solid. This means that the energy difference between the transitions occurring from a $v'$-th level of the upper electronic state to various $v''$-th levels of the ground electronic state in a solid (in our case, in a cluster) and in gas (free molecules) can be written in the following way:

$$\Delta E^{v'}(v'') = E_{cl}(v',v'') - E_{gas}(v',v'') = \delta E_m + \delta\varepsilon(v') - \delta\varepsilon(v''). \qquad (5)$$

Here $\delta E_m = E_{cl}(0,0) - E_{gas}(0,0)$ is the matrix shift of the 0-0 electronic-vibrational transition, $\delta\varepsilon(v')$ and $\delta\varepsilon(v'')$ are the differences in vibration energies of the excited and ground electronic states in clusters and in gas:

$$\delta\varepsilon(v') = v'[\hbar\omega'_{cl} - \hbar\omega'_{gas}] - v'(v'+1)[\hbar\omega'_{cl}x'_{cl} - \hbar\omega'_{gas}x'_{gas}]; \qquad (6)$$

$$\delta\varepsilon(v'') = v''[\hbar\omega''_{cl} - \hbar\omega''_{gas}] - v''(v''+1)[\hbar\omega''_{cl}x''_{cl} - \hbar\omega''_{gas}x''_{gas}]. \qquad (7)$$

According to Ref. [29], the frequency and anharmonicity of a free nitrogen molecule in the $X^1\Sigma_g^+$ and $w^1\Delta_u$ states are $\omega''_{gas}$ = 2358.6 cm$^{-1}$, $\omega''_{gas}x''_{gas}$ = 14.32 cm$^{-1}$ and $\omega'_{gas}$ = 1559.3 cm$^{-1}$, $\omega'_{gas}x'_{gas}$ = 11.63 cm$^{-1}$, respectively.

In Fig. 4, the identified $w^1\Delta_u \to X^1\Sigma_g^+$ emission series is shown as $\Delta E^{v'}(v'')$ for $v'$=0 and 1 (solid lines in Fig. 3 indicate the 0→$v''$ transitions of this series), its experimentally obtained frequencies are given in Table 1. The circles (●,○) show our experimental data $\Delta E^{v'}(v'') = E_{cl}^{\exp}(v',v'') - E_{gas}(v',v'')$ for argon-nitrogen clusters ($N \approx 250$ and $N \approx 400$ particles/cluster). The data $\Delta E^{v'}(v'')$ reported for the same series (v′=0, 1) in the bulk nitrogen spectra [30] are shown by squares (■,□), the dashed line refers to nitrogen molecules emitting from bulk argon-nitrogen samples (v′=0) [27]. These 'bulk' data can be fitted well by using the standard model for transitions in single $N_2$ molecules resulting in Eqs. (5)-(7). It can be seen for the bulk samples that $\Delta E^{v'}(v'')$=const for a given v′, i.e. $\delta\varepsilon(v'')$=0, which means that the vibrational frequency and anharmonicity in the ground electronic state remain almost unchanged when passing from gas to a solid, there is only a red matrix shift of the magnitude $\delta E_m \approx$ -320 cm$^{-1}$ and -520 cm$^{-1}$ for the Ar-$N_2$ and $N_2$ samples, respectively.

The cluster data are fitted by linear dependencies, which imply that taking clusters instead of a gas does not affects the anharmonicity, but increases the vibrational frequency[5] (by $\approx$45 cm$^{-1}$). Comparison between the 0→v″ and 1→v″ transitions give us the energy of a vibration quantum of the electronic excited state $w^1\Delta_u$ of about 1560 cm$^{-1}$ which is nearly the same value as that for a free molecule (1559.3 cm$^{-1}$). By taking v′=0 and v″=0 in Eq. (5), we can find the matrix shift $\delta E_m$ for the $w^1\Delta_u \to X^1\Sigma_g^+$ transition: $\delta E_m \approx$ -120 cm$^{-1}$.

The analyzed band series observed in mixed Ar-$N_2$ clusters can thus be attributed to the emission of single $N_2$ molecules in an Ar matrix.

### B. Anomalous electronic-vibrational luminescence spectrum from nitrogen clusters

---

[5] A small increase/decrease of the ground electronic state vibrational frequency can be observed in spectra of various diatomic molecules embedded in a rare-gas matrix [31].

In Fig. 3, by dashed lines are shown several bands which are located very close to the $w^1\Delta_u \to X^1\Sigma_g^+$ single-molecule series (solid lines; v′ = 0; v″ = 2, 3,…,7). Similar bands can also be discerned in the spectra of pure nitrogen clusters (Fig. 2). Figure 5 shows their energy positions as $\Delta E^{v'}(v'') = E_{cl}^{exp}(v',v'') - E_{gas}(v',v'')$ for $w^1\Delta_u$-state quantum numbers v′=0, 1 and 2 (see Eq. (5) for Ar-N$_2$ clusters with $N \approx 250$ and $N \approx 400$ (Fig. 5a) and N$_2$ clusters (Fig. 5b)).

These experimental data conform well to parabolic curves, while the data for single N$_2$ molecules were fitted by linear dependencies (Fig. 4). The curves in Figs. 5a and 5b can only be fitted well (that is with an error not exceeding a few percent, not shown in Fig. 5) by assuming the anharmonicity in the ground electronic state to be half its gas value ($\omega_{cl}'' x_{cl}'' = \frac{1}{2}\omega_{gas}'' x_{gas}''$) and leaving the vibrational frequency virtually unchanged ($\omega_{cl}'' = \omega_{gas}''$). Such a strong decrease in anharmonicity seems to be unfeasible for the ground state of a single molecule, since the dissociation energy D, given by the Morse potential as $D = \hbar \frac{\omega^2}{4\omega x}$, should become twice that of a free molecule, i.e., ≈20 eV, which is unrealistic.

Thus, it seems hardly possible to use the standard single-molecule Morse model to account for the observed anomalous electronic-vibrational series. A new approach is proposed in the next section.

### III. DISCUSSION

Let's analyze the observed anomalous series of electronic-vibrational transitions from the viewpoint of coupled anharmonic oscillators. Such an oscillating system has long been studied in the framework of local and normal oscillations in polyatomic molecules. The pioneer paper in this field was published by O. Redlich [32] as early as 1941. It dealt with a system of coupled identical diatomic oscillators, each of them being described by the Morse potential resulting in the quantization (3)-(4):

$$U(r) = D[1 - \exp(-\beta(r - r_e))]^2. \tag{8}$$

Here $D$ is the dissociation energy of a diatomic molecule and $\beta$ is the Morse parameter depending on its reduced mass $\mu$ and the vibration anharmonicity $\omega x$:

$$\beta = \omega(\mu/2D)^{1/2} = (2\mu\omega x/\hbar)^{1/2}; \qquad D = \frac{\hbar\omega^2}{4\omega x} \tag{9}$$

Redlich showed that, whatever the configuration of a system of N coupled oscillators, the symmetric stretching vibration of the system can be represented by a Morse potential:

$$U = ND[1 - \exp(-N^{-1/2}\beta\sigma)]^2, \tag{10}$$

where $\sigma$ is the symmetric coordinate. It follows from Eqs. (9) and (10) that the symmetric stretching vibration is characterized by the frequency ω of a single diatomic molecule and by the anharmonicity ω*x*, which is *N times smaller* than that of the single molecule (i.e., half that for two coupled oscillators).

Subsequently, there appeared a number of papers on coupling of two anharmonic oscillators. One of them [33] demonstrated that for two identical oscillators (for example, two identical bonds in a triatomic molecule) the Morse potential along the symmetric cut of the potential surface is given by:

$$U(\sigma) = 2D\left(1 - \exp\left[-\frac{\beta}{\sqrt{2}}\sigma\right]\right)^2 \qquad (11)$$

with the energies

$$\varepsilon(n) = \hbar\omega(n + 1/2) - \frac{\hbar\omega x}{2}(n + 1/2)^2, \qquad n = 0, 1, 2... \qquad (12)$$

This energy spectrum is analogous to that of a single anharmonic oscillator (see Eqs. (3) and (4)) with the anharmonicity reduced in half. This is exactly the quantization spectrum of the ground electronic state, mentioned in the previous section, which allows us to fit well the anomalous electron-vibration series in Figs. 5a and 5b.

In Refs. [34,35], the energy of two coupled Morse oscillators, each having the same vibration spectrum (4), is given by the *rotor representation*:

$$\varepsilon(m,r) = \hbar\omega(m+1)\left[1 - \frac{x}{2}(m+1)\right] - \frac{\hbar\omega x}{2}r^2 + \gamma V(m,r), \qquad (13)$$

where $n_1$ and $n_2$ are the quantum numbers of the two coupled oscillators related through the parameters $m = n_1 + n_2$ and $r = n_1 - n_2$, while $\gamma$ is the interaction constant. The minimum value of the rotor term ($r \approx 0$) corresponds to the normal mode state of the system, while non-zero values refer to local modes. In the pure normal mode limit, $r = 0$, $V(m) = m+1$ and $\gamma > \gamma_c$, the critical interaction constant being $\gamma_c = [\hbar\omega - (\hbar^2\omega^2 - 6\hbar\omega x\varepsilon)^{0.5}]/3$.

Consider the experimental data of Fig. 5 from this point of view. The solid lines are the calculated dependencies $\Delta E^{v'}(v'') = E_{coupled}(v',v'') - E_{single}(v',v'') = \delta E_m + \delta\varepsilon\ (v') - \delta\varepsilon^{theory}(v'')$, where the sum $\delta E_m + \delta\varepsilon\ (v')$ is a fitting parameter and the ground-state value $\delta\varepsilon^{theory}(v'')$ is taken as the difference between Eq. (13) for two coupled oscillators and Eq. (4) for a single oscillator. For a given value $m$ ($m = 0, 1, 2…$), we took the minimum value of $|r|$ (0 or 1) and chose the quantum numbers of a free molecule $v''$ equal to $m$: $v'' = m$. We also neglected the contribution of the oscillator interaction energy in Eq. (13), assuming that

$$\gamma V(m,r) << \hbar\omega\frac{x}{2}(m+1)^2: \quad \gamma << \hbar\omega\frac{x}{2}(m+1) \approx 80 \text{ cm}^{-1} \text{ for } m=10 \ (r=0).$$ It follows from Fig. 5 that if $v'=0$, then $\delta E_m + \delta\varepsilon\ (v')$ is -310 cm$^{-1}$ for the mixed argon-nitrogen clusters (Fig. 5a) and -370 cm$^{-1}$ for the nitrogen clusters (Fig. 5b); for $v'=1$, it is equal to -170 and -190 cm$^{-1}$, respectively; and for $v'=2$, $\delta E_m + \delta\varepsilon\ (v') = $ -10 cm$^{-1}$ (nitrogen clusters).

The experimental data of these five 'anomalous' series can be well described by a model of two weakly coupled oscillators with the vibrational frequency equal to that of a free molecule (2359 cm$^{-1}$) and the anharmonicity decreased down to 7.15 cm$^{-1}$, i.e., being half the value for a free molecule[6]. By using Eq. (6) and assuming that a molecule in the excited electronic state $w^1\Delta_u$ is a single oscillator with the anharmonicity of a free molecule, we can find the matrix shifts $\delta E_m = \delta E_m + \delta\varepsilon\ (v' = 0)$: $\delta E_m = $ -310 cm$^{-1}$ for Ar-N$_2$ clusters and -370 cm$^{-1}$ for N$_2$ clusters, and from $\delta\varepsilon\ (v' \neq 0)$ the change in the vibrational frequency of the electronic excited state. The vibrational frequency in the $w^1\Delta_u$ state turns out to be increased by 140-180 cm$^{-1}$ with respect to its gas value of 1559.3 cm$^{-1}$. We would like to note that for single molecules (Fig. 4) the vibrational frequency in this state remains virtually unchanged, as has already been mentioned above. The changes in the matrix shift and excited-state vibrational frequency of the emitting center in a system of two coupled oscillators with respect to a single oscillator in Ar-N$_2$ clusters reflects the change in the interaction energy between the radiating centers in the excited and ground states and their environment.

---

[6] If we change the vibration frequency by ±1% and the anharmonicity by ±20% (slightly correcting the matrix shift value), the model curves still fit the experimental data well.

The observed transition energies and the corresponding values of *m* and *r* are given in Table 2 for mixed argon-nitrogen clusters and in Table 3 for nitrogen clusters. It should be noted that for $m=10$ the critical value $\gamma_c$ is about 450 cm$^{-1}$ and $\gamma < \gamma_c$, therefore, the system of two weakly coupled oscillators in the clusters is likely to be in the regime of local modes interacting through exchange between the $N_2$ bonds[7].

Energy spectra of systems with a greater number of coupled oscillators can be estimated by using Eqs. (10) and (12). The dotted line in Fig. 5b shows how the $v'=0$ electron-vibration series should look for three coupled $N_2$ molecules. We can see that the energy positions of the three-oscillator emission bands can differ considerably from those of single molecules and of two coupled oscillators.

The curves for pure nitrogen clusters (Fig. 5b) suggest that in the ground electronic-vibrational state nitrogen molecules are coupled in pairs to form $(N_2)_2$ dimers with a Morse potential (11) giving a dissociation energy which is twice that inferred from the $N_2$ monomer potential and an anharmonicity parameter β which is $\sqrt{2}$ times smaller than that of a monomer. Mixed argon-nitrogen clusters, however, seem to have both paired and unpaired molecules. This raises the question about the bond type and the channels of such dimers' formation.

An $(N_2)_2$ dimer was probably first detected in gaseous nitrogen by mass spectrometry [12] and spectroscopically from infrared collision-induced absorption spectra [13]. The authors of Ref. [13] believed the observed $(N_2)_2$ species to be a "T"-shaped van der Waals dimer. According to Ref. [37], the T-conformation with the well depth of 13.3 meV (107.14 cm$^{-1}$) and equilibrium distance of 4.03 Å (separating the centers of mass of both monomers) appears to be the most stable among the five analyzed $N_2$-$N_2$ configurations[8]. Later, a $(N_2)_2$ dimer was observed in molecular gas beams [4,5], it was reported that nearly 1% of nitrogen in the beam can be dimerized under certain conditions [4].

---

[7] For low values of *m*, when the vibrational energy and the critical values of coupling are small, the system can enter the normal mode regime [35,36].
[8] In Ref. [38], the interaction potential successfully used to describe the T-shaped geometry was applied to the tetragonal (γ) phase of solid nitrogen occurring at a pressure of about 4000 atm.

We can assume three ways of dimer formation in clusters: (i) direct dimer condensation from the gaseous phase in a supersonic jet; (ii) formation of dimers through electronic excitation of $N_2$ molecules, and (iii) formation of $(N_2)_2$ in the process of growth of icosahedral clusters from liquid nanodroplets. We find the first scenario to be unlikely since, as far as we know, large clusters are formed from liquid nanodroplets which are 'heated up' by condensation of the substance and the heat should probably result in disruption of the van der Waals clusters. The second scenario implies that the cluster may originally consist of only $N_2$ molecules. By electronically exciting it, we are able to create $N_2^*$ molecules which can form $N_2^* N_2$ van der Waals complexes. In this case, an electronic transition may occur to the levels of the van der Waals complex $(N_2)_2$. These dimers can be accumulated in clusters, provided that the cluster temperature is smaller than the dimer dissociation energy. They have not, however, been observed upon excitation of bulk solid nitrogen samples. It seems likely that the third scenario can take place due to the multilayer icosahedral structure of the clusters which is not typical of bulk samples.

One of the essential differences between a multilayer icosahedral cluster and a bulk crystalline solid is that the cluster core is compressed rather strongly [39]. For example, the two inner spheres of a 3-layer icosahedral cluster ($N = 147$ particles) are closer to each other by approximately 3-5% in the radial direction in comparison with the distance between the nearest-neighbor atoms in a bulk Lennard-Jones crystal[9]. This may be responsible for the strong intensity of the observed $w^1\Delta_u \rightarrow X^1\Sigma_g^+$ transition and for the zero intensity (within the experimental error) of the $a'^1\Sigma_u^- \rightarrow X^1\Sigma_g^+$ and $a^1\Pi_g \rightarrow X^1\Sigma_g^+$ transitions in $N_2$ and Ar-$N_2$ clusters [27]. Pressure applied to solid nitrogen is known to strongly increase the oscillator strength of the $w^1\Delta_u \rightarrow X^1\Sigma_g^+$ ($\tau \approx 5\cdot10^{-3}$ s), $a'^1\Sigma_u^- \rightarrow X^1\Sigma_g^+$ ($\tau \approx 0.5$ s), and $a^1\Pi_g \rightarrow X^1\Sigma_g^+$ ($10^{-4}$ s) transitions, which lie in our frequency range and are forbidden in gas-phase spectra. The $w^1\Delta_u \to X^1\Sigma_g^+$ transition is intensified most drastically (quadratic dependence on pressure) [41], while oscillator strengths of the other transitions grow more slowly (linearly) with pressure. We can estimate the effective pressure produced by the additional 'inner' compression: it turns out to be equal to 1000-2000 atm if we take the compressibility of nitrogen being $K \approx 7 \cdot 10^{-5}$ atm$^{-1}$ at 40 K. The compression of an icosahedron's core may favor formation and accumulation of (N$_2$)$_2$ dimers.

---

[9] Compression becomes more pronounced in greater icosahedral clusters [39]. In Morse-potential clusters, surface layers are compressed, too [40].

The identified transitions are indicated by arrows in Fig. 2. According to our estimations, some of the other transitions visible in the spectrum can be assigned to the emission of $N_2^+$ ions and transitions to lower energy levels in $(N_2^+)_2$ ionic complexes. But this is only a preliminary result which is to be investigated in our future studies.

## IV. CONCLUSIONS

In the present experiments, supersonic jet spectroscopy was employed to study cathodoluminescence spectra of nitrogen clusters (containing nearly 100 molecules per cluster) and argon-nitrogen clusters (250 and 400 particles per cluster). The spectra were studied in a wide frequency range from 45000 to 73000 cm$^{-1}$ (5.6-9.1 eV). An important aspect of our experiments was the fact that the cluster beams contained either multilayer icosahedral or polytetrahedral clusters, or their mixture. In the case of mixed Ar-N$_2$ clusters, $w^1\Delta_u \to X^1\Sigma_g^+$ transitions to various vibrational levels of the ground $X^1\Sigma_g^+$ state were observed from single N$_2$ molecules in an Ar environment. Close to these emission bands, well-known for bulk samples, was detected a number of features that were never observed earlier. The latter bands made up the identified part of the emission spectra of N$_2$ clusters.

Our analysis showed that this 'anomalous' spectrum can be explained by the presence in a cluster of van der Waals (N$_2$)$_2$ dimers, which earlier were only observed in nitrogen gas and molecular beams. In the mixed clusters, the emission of (N$_2$)$_2$ dimers was accompanied by emission of single $N_2$ molecules. We can suppose that the formation of $(N_2)_2$ dimers is related to the icosahedral structures present in our cluster beams.

The results obtained can be of interest from the viewpoint of producing polymeric single-bonded nitrogen, a promising high-energy-density material, since the $(N_2)_2$ dimers can be considered a starting species for its synthesis.

## ACKNOWLEDGMENTS

The authors are grateful to Dr. S. I. Kovalenko and O. G. Danylchenko for fruitful discussions on cluster structure and cluster formation in supersonic jets.

**Figure captions**

Fig. 1. (a) Cathodoluminescence spectrum from free clusters of pure nitrogen ($N \approx 100$ molecules/cluster). (b) Cathodoluminescence spectrum from a 100-μ-thick polycrystalline sample of nitrogen in the frequency range of $w^1\Delta_u(v'=0) \rightarrow X^1\Sigma_g^+(v'')$ transitions according to Ref. [26].

Fig. 2. Detailed emission spectra from pure $N_2$ clusters ($N \approx 100$ molecules/cluster) in three frequency ranges (45000-73000 cm$^{-1}$). Arrows show the new transition series, which is not observed in the spectra of bulk samples (see text for details). Emission lines of neutral and ionized atomic nitrogen are marked as N and N$^+$.

Fig. 3. Fragment of the emission spectrum from Ar-$N_2$ clusters ($N \approx 400$ particles/cluster, $N_2$ concentration is $\approx 5\%$) with the $w^1\Delta_u \rightarrow X^1\Sigma_g^+$ electronic-vibrational transitions (indicated by vertical solid lines; v′ = 0; v″ = 2, 3,…,7) in $N_2$ molecules isolated in an Ar matrix. By dashed lines are shown several bands not observed in the spectra of bulk samples.

Fig. 4. Experimental $\Delta E^{v'}(v'') = E_{cl}^{\exp}(v',v'') - E_{gas}(v',v'')$ data for the $w^1\Delta_u \rightarrow X^1\Sigma_g^+$ series from Ar-$N_2$ clusters (●: v′=0; ○: v′=1), bulk $N_2$ samples [30] (■: v′=0; □: v′=1), and bulk Ar-$N_2$ samples with $C_{N_2} < 1$ mol.% [27] (upper dashed line). Solid line represents emission series calculated for a single $N_2$ molecule in cluster.

Fig. 5. (a) Energy positions of the experimentally observed 'anomalous' band series from Ar-$N_2$ clusters (see Table 2) after subtraction of $w^1\Delta_u \rightarrow X^1\Sigma_g^+$ gas values $E_{gas}(v',v'')$ for v′=0 and 1. Solid lines with stars are $\Delta E^{v'}(v'') = E_{coupled}(v', v'' = n_1 + n_2) - E_{single}(v',v'')$ dependencies calculated for various values $n_1$ and $n_2$ of two coupled oscillators (see Eq. (13)). (b) Energy positions of the experimentally observed 'anomalous' band series from $N_2$ clusters (see Table 3) after subtraction of $w^1\Delta_u \to X^1\Sigma_g^+$ gas values $E_{gas}(v',v'')$ for v'=0, 1, and 2. Solid lines with stars are $\Delta E^{v'}(v'') = E_{coupled}(v',v''=n_1+n_2) - E_{single}(v',v'')$ dependencies calculated for various values $n_1$ and $n_2$ of two coupled oscillators (see Eq. (13)). Dotted line is a calculated dependence of the v'=0 electron-vibration series for three coupled $N_2$ molecules.

## Tables

TABLE 1. Electron-vibration transitions $w^1\Delta_u(v') \rightarrow X^1\Sigma_g^+(v'')$ in single N$_2$ molecules from mixed Ar-N$_2$ clusters.

| Transition energy, cm$^{-1}$ | v′ → v″ |
|---|---|
| not observed | 0 → 0 |
| not observed | 0 → 1 |
| 66818 | 0 → 2 |
| 64501 | 0 → 3 |
| 62322 | 0 → 4 |
| 59966 | 0 → 5 |
| 57743 | 0 → 6 |
| 55560 | 0 → 7 |
|  |  |
| not observed | 1 → 0 |
| not observed | 1 → 1 |
| not observed | 1 → 2 |
| 66056 | 1 → 3 |
| 63775 | 1 → 4 |
| not observed | 1 → 5 |
| 59313 | 1 → 6 |
| 57120 | 1 → 7 |
| 54936 | 1 → 8 |
| 52751 | 1 → 9 |

Table 2. Electron-vibration transitions from the upper $w^1\Delta_u$ (v′=0, 1) state to the ground state of two coupled N$_2$($X^1\Sigma_g^+$) + N$_2$($X^1\Sigma_g^+$) molecules in mixed Ar-N$_2$ clusters.

| Transition energy, cm$^{-1}$ | v′ | m | \|r\| | $n_1, n_2$ |
|---|---|---|---|---|
| not observed |  | 0 | 0 | 0,0 |

| Transition energy, cm$^{-1}$ | v′ | m | \|r\| | n$_1$,n$_2$ |
|---|---|---|---|---|
| 69085 | 0 | 1 | 1 | 1,0 |
| not observed | | 2 | 0 | 1,1 |
| 64380 | | 3 | 1 | 2,1 |
| 62177 | | 4 | 0 | 2,2 |
| 59831 | | 5 | 1 | 3,2 |
| 57559 | | 6 | 0 | 3,3 |
| 55284 | | 7 | 1 | 4,3 |
| 53108 | | 8 | 0 | 4,4 |
| not observed | 1 | 0 | 0 | 0,0 |
| 70756 | | 1 | 1 | 1,0 |
| 68280 | | 2 | 0 | 1,1 |
| 66183 | | 3 | 1 | 2,1 |
| 63579 | | 4 | 0 | 2,2 |
| 61610 | | 5 | 1 | 3,2 |
| 59279 | | 6 | 0 | 3,3 |
| 56994 | | 7 | 1 | 4,3 |
| 54791 | | 8 | 0 | 4,4 |
| 52617 | | 9 | 1 | 5,4 |

TABLE 3. Electron-vibration transitions from the upper $w^1\Delta_u$(v′=0, 1, 2) state to the ground state of two coupled N$_2$($X^1\Sigma_g^+$) + N$_2$($X^1\Sigma_g^+$) molecules in pure N$_2$ clusters.

| Transition energy, cm$^{-1}$ | v′ | m | \|r\| | n$_1$,n$_2$ |
|---|---|---|---|---|
| not observed | 0 | 0 | 0 | 0,0 |
| 68984 | | 1 | 1 | 1,0 |
| 66753 | | 2 | 0 | 1,1 |
| 64340 | | 3 | 1 | 2,1 |
| 62132 | | 4 | 0 | 2,2 |
| 59793 | | 5 | 1 | 3,2 |
| 57495 | | 6 | 0 | 3,3 |
| 55274 | | 7 | 1 | 4,3 |
| not observed | | 0 | 0 | 0,0 |
| 70781 | | 1 | 1 | 1,0 |
| 68299 | | 2 | 0 | 1,1 |
| 66154 | | 3 | 1 | 2,1 |
| not observed | | 4 | 0 | 2,2 |
| 61540 | | 5 | 1 | 3,2 |
| 59244 | 1 | 6 | 0 | 3,3 |
| 56956 | | 7 | 1 | 4,3 |
| 54721 | | 8 | 0 | 4,4 |
| 52508 | | 9 | 1 | 5,4 |
| 50283 | | 10 | 0 | 5,5 |
| not observed | | 11 | 1 | 6,5 |
| 45982 | | 12 | 0 | 6,6 |
| not observed | | 0 | 0 | 0,0 |
| 72336 | | 1 | 1 | 1,0 |
| 70108 | | 2 | 0 | 1,1 |
| not observed | | 3 | 1 | 2,1 |
| 65524 | | 4 | 0 | 2,2 |
| 63159 | | 5 | 1 | 3,2 |
| not observed | | 6 | 0 | 3,3 |
| 58705 | 2 | 7 | 1 | 4,3 |
| 56426 | | 8 | 0 | 4,4 |
| 54204 | | 9 | 1 | 5,4 |
| 52006 | | 10 | 0 | 5,5 |
| 49846 | | 11 | 1 | 6,5 |
| 47682 | | 12 | 0 | 6,6 |
| 45516 | | 13 | 1 | 7,6 |

# References

[1] J. Farges, M.-F. de Feraudy, B. Raoult, and G. Torchet, Adv. Chem. Phys. **70**, 45 (1988).

[2] A. G. Danil'chenko, S. I. Kovalenko, and V. N. Samovarov, Low Temp. Phys. **31**, 979 (2005).

[3] H.-D. Barth, F. Huisken, and A. A. Ilyukhin, Appl. Phys. B **52**, 84 (1991).

[4] F. Carnovale, J. B. Peel, and R. G. Rothwell, J. Chem. Phys. **88**, 642 (1988).

[5] K. Norwood, G. Luo, and C. Y. Ng, J. Chem. Phys. **91**, 849 (1989).

[6] E. E. Rennie and P. M. Mayer, J. Chem. Phys. **120**, 10561 (2004).

[7] F. Calvo, G. Torchet, and M.-F. de Feraudy, J. Chem. Phys. **111**, 4650 (1999); O. G. Danylchenko, S. I. Kovalenko, and V. N. Samovarov, Low Temp. Phys. **32**, 1182 (2006).

[8] O. G. Danylchenko, S. I. Kovalenko, and V. N. Samovarov, Low Temp. Phys. **33**, 1043 (2007).

[9] J. Stapelfeldt, J. Wörmer, and T. Möller, Phys. Rev. Lett. **62**, 98 (1989); J. Wörmer, M. Joppien, G. Zimmerer, and T. Möller, Phys. Rev. Lett. **67**, 2053 (1991); M. Runne, J. Becker, W. Laasch, D. Varding, G. Zimmerer, M. Liu, and R. E. Johnson, Nucl. Instrum. Methods B **82**, 301 (1993); R. von Pietrowski, K. von Haeften, T. Laarmann, T. Möller, L. Museur, and A. V. Kanaev, Eur. Phys. J. D **38**, 323 (2006).

[10] R. Müller, M. Joppien, and T. Möller, Z. Phys. D **26**, 370 (1993); E. T. Verkhovtseva, E. A. Bondarenko, and Yu. S. Doronin, Low Temp. Phys. **30**, 34 (2004); Yu. S. Doronin, V. N. Samovarov, and E. A. Bondarenko, Low Temp. Phys. **32**, 251 (2006).

[11] O. G. Danylchenko, Yu. S. Doronin, S. I. Kovalenko, M. Yu. Libin, V. N. Samovarov, and V. L. Vakula, Phys. Rev. A **76**, 043202 (2007); V. L. Vakula, O. G. Danylchenko, Yu. S. Doronin, S. I. Kovalenko, M. Yu. Libin, and V. N. Samovarov, Low Temp. Phys. **35**, 944 (2009).

[12] E. J. Robbins and R. E. Leckenby, Nature **206**, 1253 (1965).

[13] Ch. A. Long, G. Henderson, and G. E. Ewing, Chem. Phys. **2**, 485 (1973).

[14] F. Cacace, G. de Petris, and A. Troiani, Science **295**, 480 (2002).

[15] M. T. Nguyen, Coord. Chem. Rev. **244**, 93 (2003).

[16] A. F. Goncharov, E. Gregoryanz, H.-k. Mao, Zh. Liu, and R. J. Hemley, Phys. Rev. Lett. **85**, 1262 (2000).

[17] M. Popov, Phys. Lett. A **334**, 317 (2005).

[18] A. Trojan, M. I. Eremets, S. A. Medvedev, A. G. Gavriliuk, and V. B. Prakapenka, Appl. Phys. Lett. **93**, 091907 (2008).

[19]. Th. G. Manning and Z. Iqbal, 1st Annual National Capital Region Energetics Symposium, La Plata, Maryland, 2009 (http://www.etcmd.org/conference-docs/presentations/manning-polymeric-nitrogen.pdf); H. Abou-Rachid, A. Hu, D. Arato, X. Sun, and S. Desilets, Int. J. Energetic Materials Chem Prop. **7**, 359 (2008).

[20] J. W. Hewage and F. G. Amar, J. Chem. Phys. **119**, 9021 (2003).

[21] R. A. Smith, T. Ditmire, and J. W. G. Tisch, Rev. Sci. Instrum. **69**, 3798 (1998).

[22] G. N. Makarov, Phys. Usp. **49**, 117 (2006).

[23] A. Mackay, Acta Crystallogr. **15**, 916 (1962).

[24] O. G. Danylchenko, Yu. S. Doronin, S. I. Kovalenko, and V. N. Samovarov, JETP Lett. **84**, 324 (2006); Yu. S. Doronin and V. N. Samovarov, Opt. Spectrosc. **102**, 906 (2007).

[25] E. V. Gnatchenko, A. N. Nechay, V. N. Samovarov, and A. A. Tkachenko, Phys. Rev. **82**, 012702 (2010).

[26] Yu. B. Poltoratskii, V. M. Stepanenko, and I. Ya. Fugol', Sov. J. Low Temp. Phys. **7**, 60 (1981).

[27] I. Ya. Fugol' and Yu. B. Poltoratski, Solid State Commun. **30**, 497 (1979); I.Ya. Fugol', Yu.B. Poltoratskii, and E.V. Savchenko, JETP Lett. **24**, 1 (1976).

[28] M. Peyron and H.P. Broida, J. Chem. Phys. **30**, 139 (1959).

[29] K. P. Huber and G. Herzberg, *Molecular Spectra and Molecular Structure* (Van Nostrand Reinhold Co., New York, 1979).

[30] V. M. Stepanenko, Ph.D. thesis, B. Verkin Institute for Low Temperature Physics and Engineering, 1984.

[31] M. E. Jacox, Chem. Phys. **189**, 149 (1994).

[32] O. Redlich, J. Chem. Phys. **9**, 298 (1941).

[33] E. J. Heller and W. M. Gelbart, J. Chem Phys. **73**, 626 (1980).

[34] E. L. Sibert, J. T. Hynes, and W. P. Reinhardt, J. Chem. Phys. **77**, 3595 (1982).

[35] D. C. Rouben and G. S. Ezra, J. Chem. Phys. **103**, 1375 (1995).

[36] M. P. Jacobson, R. J. Silbey, and R. W. Field, J. Chem. Phys. **110**, 845 (1999).

[37] V. Aquilanti, M. Bartolomei, D. Cappelletti, E. Carmona-Novillo, and F. Pirani, J. Chem. Phys. **117**, 615 (2002).

[38] J. C. Raich and N. S. Gillis, J. Chem. Phys. **66**, 846 (1977).

[39] J. Farges, M. F. de Feraudy, B. Raoult, and G. Torchet, Adv. Chem. Phys. **70**, 45 (1988).

[40] M. R. Hoare and P. Pal, Nature Phys. Sci. **236**, 35 (1972).

[41] A. Lofthus and P. H. Krupenie, J. Phys. Chem. Ref. Data **6**, 113 (1977).

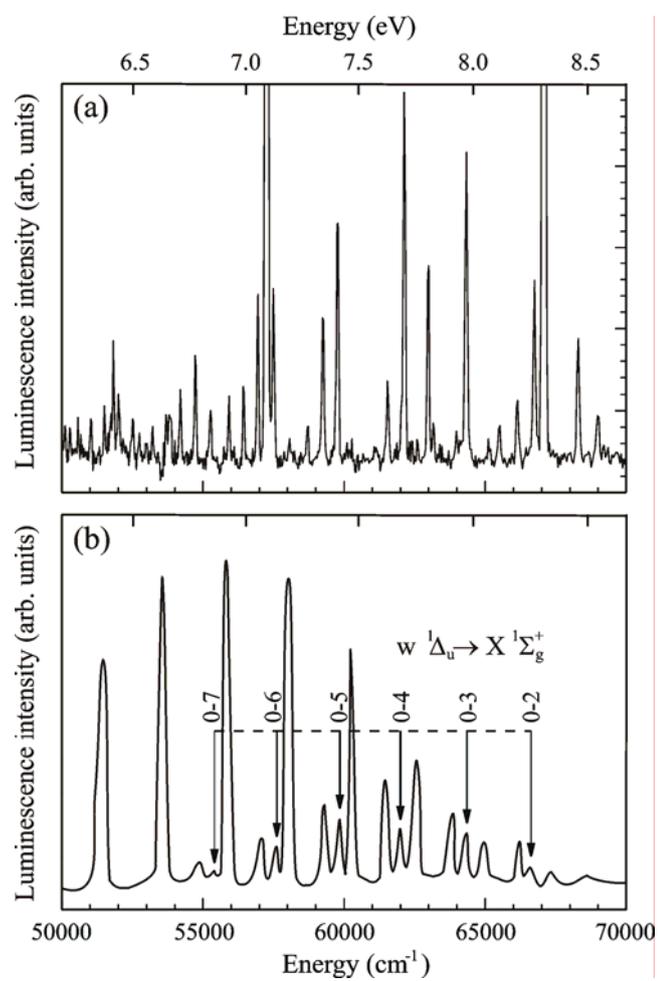

Fig. 1

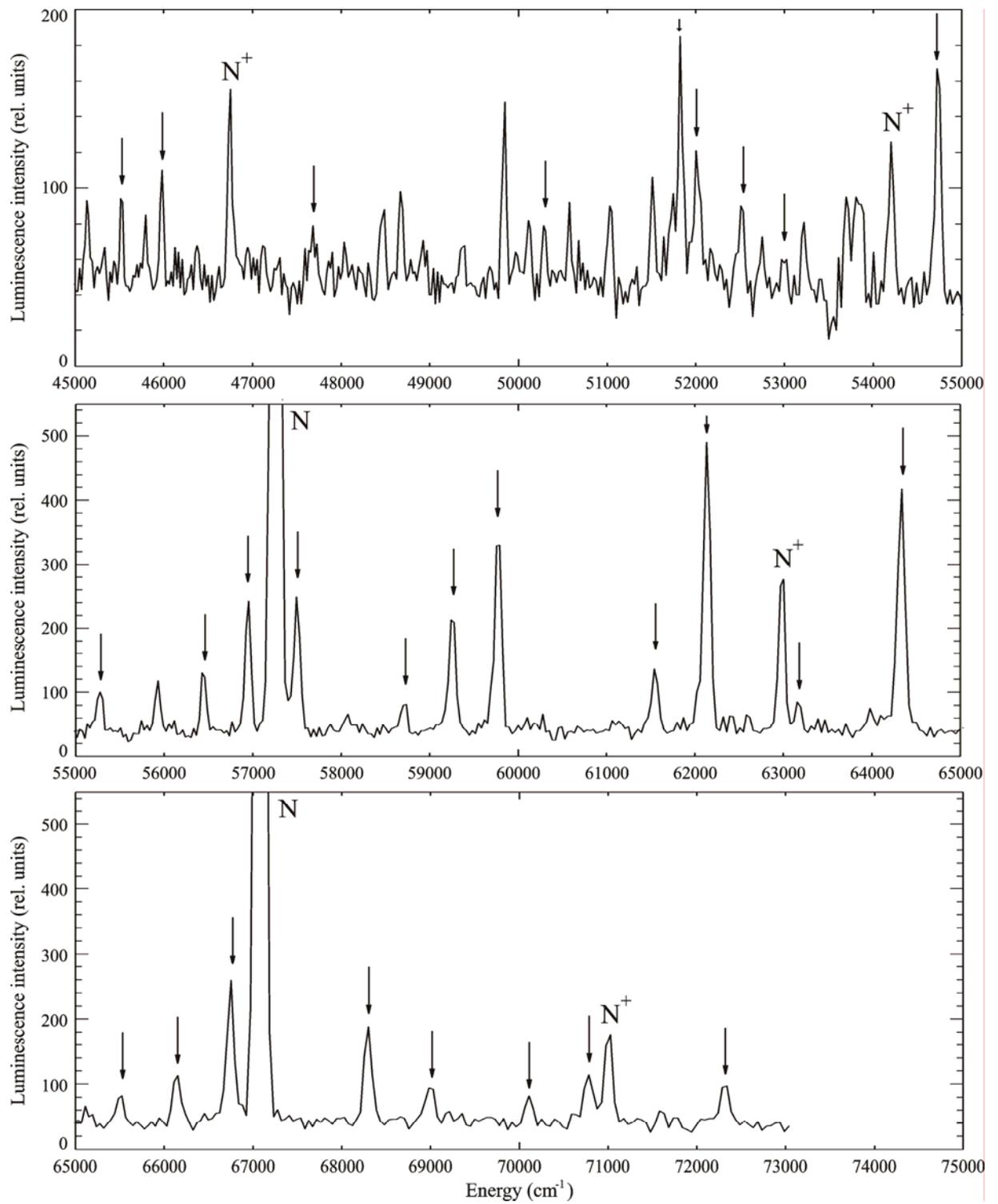

Fig. 2

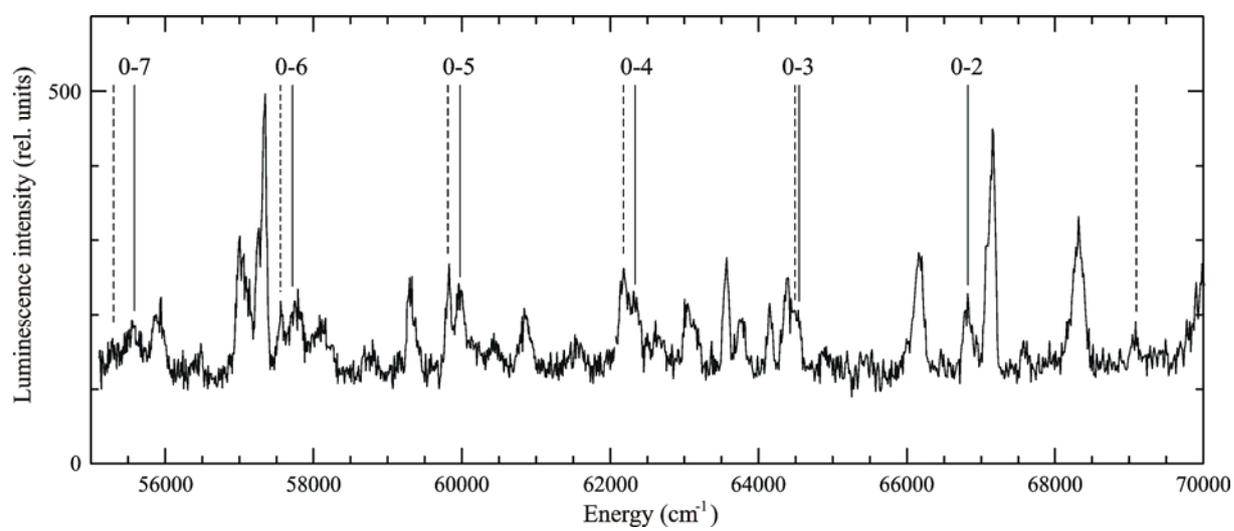

Fig. 3

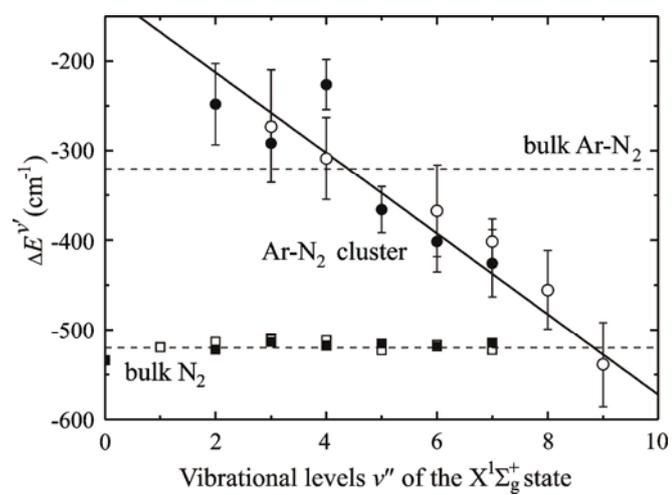

Fig. 4

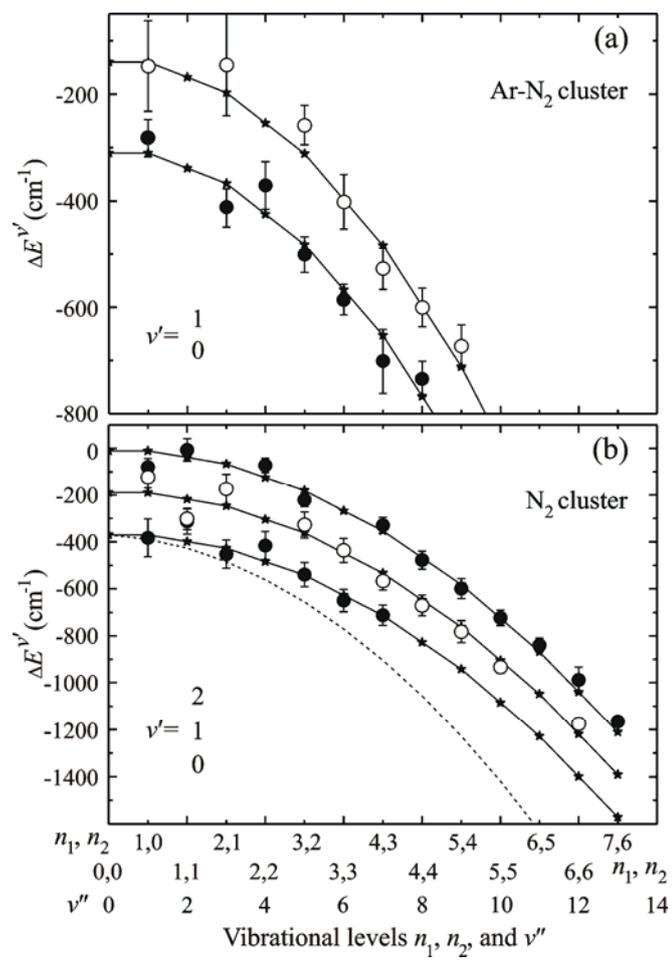

Fig. 5